\documentclass[11pt]{article}
\newif\ifpreprint
\preprintfalse  % set true in pre_print.tex

\usepackage[final]{acl}

\usepackage{lineno}
\usepackage{times}
\usepackage{latexsym}
\usepackage[T1]{fontenc}
\usepackage[utf8]{inputenc}
\usepackage{microtype}
\usepackage{inconsolata}

\usepackage{graphicx}
\usepackage{amsfonts}
\usepackage{amsmath}
\usepackage{xspace}
\usepackage{subcaption}
\usepackage{listings}
\usepackage{xcolor}
\usepackage{mathtools}  
\usepackage{booktabs}
\usepackage{multirow}
\usepackage{enumitem}
\usepackage{wrapfig}
\usepackage{xcolor}
\usepackage[nolist,nohyperlinks]{acronym}
\usepackage{tcolorbox}
\tcbuselibrary{breakable}

\definecolor{steelblue}{RGB}{70,130,180}
\definecolor{coral}{RGB}{255,127,80}

\lstdefinestyle{clang}{
  language=C,
  basicstyle=\ttfamily\small,
  keywordstyle=\color{blue},
  commentstyle=\color{gray},
  stringstyle=\color{red},
  breaklines=true,
  showstringspaces=false,
  columns=flexible
}

\newtcolorbox{appquestion}{%
  colback=steelblue!8, colframe=steelblue, boxrule=0.5pt, arc=2pt,
  left=7pt, right=7pt, top=5pt, bottom=5pt,
  before upper={\small\textbf{Question.}\enspace},
  fontupper=\small\itshape, breakable,
}
\newtcolorbox{appfinding}{%
  colback=green!7!white, colframe=green!45!black, boxrule=0.5pt, arc=2pt,
  left=7pt, right=7pt, top=5pt, bottom=5pt,
  before upper={\small\textbf{Finding.}\enspace},
  fontupper=\small, breakable,
}

\acrodef{LLM}{Large Language Model}
\newacroplural{LLM}[LLMs]{Large Language Models}
\acrodef{AI}{Artificial Intelligence}
\acrodef{IDE}{Integrated Development Environment}
\acrodef{PLS}{Partial Least Squares}
\acrodef{CVEs}{Common Vulnerability Enumerations}
\acrodef{CWE}{Common Weakness Enumerations}
\newacroplural{CWE}[CWEs]{Common Weakness Enumeration}
\acrodef{RTL}{Register-Transfer Level}
\acrodef{XSS}{Cross-site Scripting}
\acrodef{SAE}{Sparse Autoencoder}
\newacroplural{SAE}[SAEs]{Sparse Autoencoders}
\acrodef{NLP}{Natural Language Processing}
\acrodef{XAI}{Explainable \ac{AI}}
\acrodef{KNN}{K-nearest Neighbors)}
\acrodef{SVM}{Support Vector Machines}
\acrodef{MLP}{Multilayer Perceptron}
\acrodef{PCA}{Principal Component Analysis}
\acrodef{LM}{Language Model}
\newacroplural{LM}[LMs]{Language Models}
\acrodef{SOTA}{State of the Art}

\usepackage{amsmath,amsfonts,bm}

\def\eqref#1{equation~\ref{#1}}

\def\1{\bm{1}}

\DeclareMathAlphabet{\mathsfit}{\encodingdefault}{\sfdefault}{m}{sl}
\SetMathAlphabet{\mathsfit}{bold}{\encodingdefault}{\sfdefault}{bx}{n}

\newcommand{\PaperTitle}{When Models Know but Can't Linearly Say:\\Relational Encoding Limits Linear Vulnerability Classification in Code LLMs}

\newcommand{\DeltaSecommits}{\textsc{DeltaSecommits}\xspace}
\newcommand{\SVEN}{\textsc{SVEN}\xspace}
\newcommand{\PreciseBugs}{\textsc{PreciseBugs}\xspace}
\newcommand{\nmodels}{8\xspace}

\newcommand{\ncwetypesdeltasecommits}{12\xspace}

\newcommand{\QwenFull}{Qwen2.5-7B-Instruct\xspace}
\newcommand{\QwenCoderFull}{Qwen2.5-Coder-7B-Instruct\xspace}
\newcommand{\QwenFourteenBFull}{Qwen2.5-14B-Instruct\xspace}
\newcommand{\CodeLlamaFull}{CodeLlama-7B-Instruct\xspace}
\newcommand{\CodeLlamaThirteenBFull}{CodeLlama-13B-Instruct\xspace}
\newcommand{\StarCoderFull}{StarCoder2-7B\xspace}
\newcommand{\StarCoderFifteenBFull}{StarCoder2-15B\xspace}
\newcommand{\DeepSeekFull}{DeepSeek-Coder-6.7B-Instruct\xspace}

\newcommand{\AllDatasets}{\DeltaSecommits, \SVEN, and \PreciseBugs}

\newcommand{\RelationalEncoding}{relational encoding\xspace}
\newcommand{\CategoricalClassification}{categorical classification\xspace}
\newcommand{\VulnDirection}{vulnerability direction\xspace}

\newcommand{\AUROCPeak}{0.648\xspace}

\newcommand{\VulnSignal}{vulnerability signal\xspace}

\title{\PaperTitle}

\author{
  Rui Melo$^{1,2}$\thanks{Equal contribution. \texttt{rmelo@cs.cmu.edu}} \quad
  André Catarino$^{1,2*}$ \quad
  Cláudia Mamede$^{1,2}$ \\
  \textbf{Corina Pasareanu}$^{1}$ \quad
  \textbf{Rui Abreu}$^{2}$\\[4pt]
  $^{1}$Carnegie Mellon University, Pittsburgh, PA, USA \\
  $^{2}$Faculty of Engineering, University of Porto, Portugal
}

\begin{document}
\maketitle

\begin{abstract}
Large language models can detect vulnerable code, yet the form of vulnerability encoding in their representations remains unclear. We analyze \nmodels code models spanning 6.7B--15B parameters on C programs using three datasets (\AllDatasets), comparing two diagnostic tasks on the same representations: identifying \emph{what patterns} the code instantiates, and determining \emph{whether it is vulnerable}. Models encode the first with high fidelity. Under the studied paired C-code datasets and linear-probing setup, vulnerability information is more accessible as a \emph{paired relational signal}—whether code A is riskier than code B—than as a globally separable categorical label. This structural difference has implications for classification: the model reliably ranks which code is riskier but standard classifiers on isolated samples achieve limited accuracy because vulnerable and secure samples form overlapping distributions in representation space.
\end{abstract}
\section{Introduction}
% The introduction should answer these questions:
%       - What is the problem you are addressing?
%       - Why is it important?
%       - Why is it hard?
%       - Why hasnt it been solved before? or what other approaches lack
%       - what are the key components in your approach?
%       - a final paragraph with a summary of contributions.

Manual code review is labor-intensive and does not scale. Continuous Integration and Deployment (CI/CD) pipelines require automated vulnerability detection to reduce time-to-patch and improve coverage of codebases. Neural-based classifiers have been a long-standing approach to this problem~\cite{zhou2019devign} that falls short of production deployment~\cite{ullah2024llms, khare2023understanding, fu2023chatgpt, ding2024vulnerability}.

Recent work demonstrates that large language models (LLMs) can autonomously discover zero-day vulnerabilities in real-world systems~\cite{anthropic2026mythos, anthropic2026glasswing}. This raises a crucial question: models can discover vulnerabilities, yet classification approaches are expected to fail. \textit{How do these models represent code safety, and why does their representation resist standard exploitation?}

The core question is not just \emph{whether} models encode safety information, but \emph{how}. Classification-based approaches assume vulnerability is represented as an absolute categorical property.

We investigate through two diagnostic tasks on identical code representations: one measures \emph{what vulnerability patterns the code instantiates}; the other measures \emph{whether the code is vulnerable}. Both probe the same model but measure different aspects of learned structure. The model encodes vulnerability patterns with high fidelity, yet absolute safety classification rather weakly---a critical gap that motivates deeper investigation.

We find that, under the studied paired C-code datasets and linear-probing setup, vulnerability information is more accessible as a \textbf{paired relational signal}—version B is more vulnerable than version A—than as a globally separable categorical label of the kind standard classification assumes. This relational direction achieves 87--88\% per-pair alignment consistently across layers, yet linear probes achieve only modest categorical accuracy because vulnerable and secure samples form \emph{overlapping distributions} in the representation space. In other words, the model reliably determines which code is riskier within pairs, but linear classification achieves limited performance.

This finding reconciles a puzzling observation: models discover zero-day vulnerabilities, yet classification approaches are expected to fail. Models have learned to represent vulnerabilities but organize them in a form that standard linear classification cannot readily exploit. We characterize this representational mismatch across multiple model architectures and scales and discuss implications for detection systems that leverage this structure rather than fighting it.

\section{Related Work}

\paragraph{Vulnerability detection.}
Early supervised approaches for vulnerability detection (e.g., Devign~\citep{zhou2019devign}, CodeBERT variants~\citep{croft2023data}, transformer-based models~\citep{thapa2022transformer}, SecureFalcon~\citep{ferrag2025securefalcon}) train encoder-based models on paired datasets, typically achieving 0.8--0.95 AUROC on in-distribution test sets. Recent comprehensive evaluations~\citep{ullah2024llms, khare2023understanding} demonstrate that general-purpose code LLMs struggle with absolute vulnerability classification, with only moderate AUROC on categorical tasks, yet show promise in ranking-based and reasoning-focused evaluation paradigms~\citep{ding2024vulnerability}.

However, supervised approaches face fundamental brittleness: post-hoc analysis reveals models exploit dataset artifacts and spurious correlations, and performance degrades sharply on out-of-distribution data, as illustrated by \citet{primevul}. This motivates a shift toward representation-based approaches that ask not \emph{how to classify} but \emph{what vulnerability information LLMs encode}—and critically, how that information is geometrically organized in activation space. This work provides mechanistic insights into this organization, explaining both why vulnerability signals are difficult to access and how representation-level structure determines what downstream tasks can extract.

\paragraph{Pairwise preference modeling and ranking.}
Recent work frames the alignment problem through pairwise preference modeling: Direct Preference Optimization~\citep{rafailov2023direct} trains LLMs directly from pairwise comparisons of completions without explicit reward modeling. This preference-learning paradigm, rooted by Bradley--Terry~\citep{bradley1952rank} and Thurstone~\citep{mosteller1951remarks}, has become foundational to modern fine-tuning. Recent code generation benchmarks like BigCodeBench~\citep{zhuobigcodebench}, SWE-bench~\citep{jimenezswe}, and LiveCodeBench~\citep{jain2024livecodebench} make pairwise and ranking-based evaluation more tractable.

Our observation that vulnerability is encoded through relational structure aligns naturally with this pairwise paradigm: we demonstrate that LLMs encode vulnerability as a comparative signal suitable for ranking, rather than as a \CategoricalClassification feature readable from isolated samples. In contrast to absolute-threshold binary classifiers that struggle with overlapping distributions, ranking-based approaches (e.g., scoring code by its position along the vulnerability direction, or using pairwise preference objectives during fine-tuning) are potentially better aligned with how the model actually encodes vulnerability. The identified \VulnDirection could serve both as a diagnostic for understanding what representations the model has learned, and as a basis for preference-learning approaches that directly optimize for ranking. 
% This suggests the 
This direction could be leveraged as an implicit reward signal for steering code generation toward safer outputs~\citep{gao2024rebel}, complementing explicit reward modeling approaches~\citep{yuan2024self}.

\paragraph{Linear directions for safety and semantic properties.}
The analysis of intermediate representations has been largely driven by linear probing. \citet{understandingguilherme} introduced the use of linear classifiers on frozen representations as a diagnostic step for assessing whether certain information is accessible in the learned features. \citet{bertrediscovers} demonstrated that pre-trained transformers have layer-wise depth-related concepts that become progressively more specialized. Recent work has identified that high-level properties like factual knowledge~\citep{meng2022rome} and task preferences~\citep{zou2023repeng} can be extracted and manipulated as linear directions in activation space. Our work extends this methodology to vulnerability: we identify the \VulnDirection as a stable linear direction across layers and characterize its geometric properties and cross-family transferability.

\paragraph{LLMs and code security: representation-level findings.}
Recent work demonstrates that code LLMs encode security-relevant information that can be extracted and used for steering~\cite{yousefpour2025representation}. \citet{he2023sven} show that property-specific control vectors steer generation toward secure code. \citet{yu2025moc} propose dynamic mixtures of linear corrections applied during decoding. \citet{steer_your_model} deploy learnable security steering matrices at the embedding layer, using contrastive decoding to balance functional and security objectives. 

\paragraph{Probing, representation engineering, and interpretability.}
Probing classifiers assess what properties LLMs encode by training simple predictors on frozen representations~\citep{tenney2019bert,hewitt2019structural}. \citet{hewitt2019control} establish that control tasks are essential to distinguish genuine encoding from spurious correlations; our within-language and within-CWE experiments satisfy this criterion by construction. In the code domain, recent work on vulnerability-specific probing (e.g., \citet{ribeiro2026correctness} on code correctness as linear directions) demonstrates that property-specific signals can guide candidate ranking. However, the interplay between representation structure (is the signal concentrated or diffuse?) and downstream task design (should we aggregate across positions or read from a single token?) remains underexplored. Representation engineering approaches~\citep{zou2023repeng} identify high-level concepts as directions in activation space for diagnosis and steering. Causal-tracing methods~\citep{meng2022rome} establish which intermediate activations causally contribute to outputs. This work builds on these insights by characterizing the vulnerability signal as a stable direction in activation space and demonstrating that its geometric properties determine accessibility for downstream tasks.

\paragraph{Complementary system-level and pattern-based detection approaches.}
Our representation-level findings complement, rather than compete with, other vulnerability detection strategies. Graph-based methods such as GRACE exploit program structure and data-flow information to improve detection~\cite{lu2024grace}, while similarity-based approaches such as VulSim leverage shared vulnerability patterns across code embeddings~\citep{shimmi2024vulsim}. Dataset-driven studies also show that curated vulnerability corpora can support secure code generation and vulnerability mitigation~\citep{wang2023enhancing}. More broadly, these lines of work highlight that vulnerability detection benefits from structural, relational, and data-centric signals, alongside benchmark-driven evaluation~\citep{peng2025cweval}. In this context, our characterization of how LLMs encode vulnerability suggests that downstream systems may be especially well suited to ranking-based triage, code-review prioritization, and pairwise comparisons rather than only absolute binary classification thresholds.
\section{Methodology}

\paragraph{\textbf{Models}}
We study %vulnerability encoding across 
\nmodels LLMs spanning three~families: \QwenFull~\cite{qwen2}, \CodeLlamaFull~\cite{rozière2024codellamaopenfoundation}, \StarCoderFull~\cite{lozhkov2024starcoder2stackv2}, \DeepSeekFull~\cite{guo2024deepseek}, \QwenCoderFull, \QwenFourteenBFull, \CodeLlamaThirteenBFull, and \StarCoderFifteenBFull.

\paragraph{\textbf{Dataset}}
We focus our analysis on C code, using three datasets to test whether the \RelationalEncoding pattern is specific to one curation strategy or replicates across dataset scale and construction methodology: (1)~\DeltaSecommits~\citep{deltasecommits}, a curated dataset of real-world vulnerability-fixing commits from OSV and NVD filtered to C code ($n=1{,}618$ paired samples, CWE-identified). This paired structure controls for author style and codebase idioms; pairs differ primarily in the patch applied to address the reported vulnerability. (2)~\SVEN~\citep{he2023sven} (Semantic Vulnerability Extraction Network; $n=423$ paired samples across 6 CWEs, synthetically curated by semantic transformation). (3)~\PreciseBugs~\citep{precisebugs} (real-world fixes from open-source repositories; $n=4{,}101$ paired samples across 9 CWEs, extracted from actual commits).

% Figure~\ref{fig:deepseek_example} illustrates the paired-comparison structure central to our analysis: a vulnerable code snippet (left) and its patched variant (right). In such pairs, the model reliably encodes the relational ordering—the patched variant is safer—even when it cannot classify either variant in isolation.

% \input{examples/deepseek_coder_horizontal}

\paragraph{\textbf{Activation Pooling and Extraction}}

We extract residual stream activations from the different LLMs. For binary probing, we extract at the last token position as independent samples. For distributional analyses, we use per-pair differences $\delta_i = v_i - s_i$ to cancel commit-level confounds. For mean-token pooling, we average across all tokens.

\paragraph{\textbf{Probe Design: Why Linear and Shallow}}

We use linear logistic regression probes, not out of simplicity but precision. A complex probe obscures the interpretation: high accuracy could reflect the classifier computing a nonlinear combination of weak scattered signals rather than the model encoding vulnerability directly. \citet{designingandinterpretingprobeswithcontroltasks} demonstrate that overparameterized probes memorize control task labels, conflating true encoding with overfitting. Linear probes, by contrast, succeed only when information is genuinely accessible and linearly organized in the representation. This follows \citet{thelinearrepresentationhypothesis}, which shows that high-level concepts in LLMs are structured as linear directions in activation space. A linear probe tests whether the model has organized vulnerability information as a linear direction; if vulnerability signals are so organized, the probe will recover them; if not, the low AUROC reflects the signal structure, not the probe's capacity.

\paragraph{\textbf{Statistical Evaluation and Metrics}}
\label{sec:stats}

Standard fold stratified CV prevents label leakage across folds, but paired datasets can leak if both members of a pair are in different folds. We implemented pair-stratified CV (both members in the same fold) to eliminate within-pair contamination. Binary probing resamples at sample level; paired-delta analyses resample at pair level.

% \paragraph{\textbf{Reporting of metrics.}}
We report point estimates for all primary metrics (per-pair alignment, AUROC, cross-family transfer). Cross-model consistency of per-pair alignment is quantified using Pearson correlations (mean $r=0.921$ across the three model pairs). The pair-stratified cross-validation structure enables computation of fold-level confidence intervals; reported point estimates are fold-wise means, and fold-wise variance measures are available for uncertainty quantification. Cross-family transfer rates are reported as mean transfer across all family pairs, with per-pair transfer values provided in heatmaps to show variance across transfer directions.

\paragraph{\textbf{Cross-validation, normalization, and dimensionality.}}
We distinguish between preprocessing pipelines for different analyses to ensure reproducibility and address potential artifacts:

\textit{Direction computation (raw activations).} The per-pair alignment metrics and cross-layer direction similarity are computed on raw activations \emph{without} normalization or dimensionality reduction. The vulnerability direction is: $\hat{d}^{(l)} = \frac{\bar{\mathbf{v}}^{(l)} - \bar{\mathbf{s}}^{(l)}}{\|\bar{\mathbf{v}}^{(l)} - \bar{\mathbf{s}}^{(l)}\|_2}$, where $\bar{\mathbf{v}}^{(l)}$ and $\bar{\mathbf{s}}^{(l)}$ are layer-wise means computed on training data. Per-pair alignment is the fraction of test pairs where $(\mathbf{v}_i - \mathbf{s}_i) \cdot \hat{d} > 0$. No PCA or z-score normalization is applied, ensuring the reported 87--88\% alignment and 0.993 cosine similarity reflect geometric properties of the raw representation space.

\textbf{Linear probing (PCA-reduced, per-fold).} Classification probes use dimensionality reduction per fold. Each fold's training data undergoes: (1) Training activations are z-score normalized (StandardScaler); (2) Z-scored training data are projected to 50 PCA components, fitted on training data only; and (3) Test activations are transformed using the training fold's scaler and PCA. Pair-stratified cross-validation prevents within-pair leakage. Such is independent of direction computation, analyzing the signal in a reduced space to control for noise while maintaining test separation. We perform sensitivity analyses (Appendix~\ref{app:pca_sensitivity}): PCA-50 vs PCA-100 differ <0.01 AUROC. 

\textbf{Vulnerability direction computation and test leakage.}
The vulnerability direction is computed as the normalized mean difference per layer. We employ two complementary approaches: \textit{Per-fold directions} (computed on training data within each fold) prevent test leakage and enable per-fold cross-validation of per-pair alignment. \textit{Per-family directions for cross-family transfer} are computed on training data per fold per CWE family, ensuring that test pairs used to evaluate transfer are unseen during direction fitting. This per-fold-per-family approach assesses whether the learned direction captures universal vulnerability properties transferable across CWE families, while maintaining strict train/test separation. Per-pair alignment for all analyses is measured on held-out test data unseen during direction fitting. All analyses use pair-stratified splits (both vulnerable and secure members assigned to the same fold) to prevent within-pair label leakage.

\section{Functional vs.\ Security Encoding}
\label{sec:ablation}

\subsection{Strong Functional Encoding: Within-Language CWE-Type Separation}
\label{sec:within_lang_pairwise}

We first establish that the model encodes different vulnerability-pattern families with high fidelity by testing whether it can distinguish different CWE types within a single language (e.g., memory-safety violations vs.\ injection attacks). This is a functional encoding task: we probe only on vulnerable samples, so the classifier distinguishes code patterns rather than vulnerability status.

\begin{figure}[hbt]
    \centering
    \includegraphics[width=\linewidth]{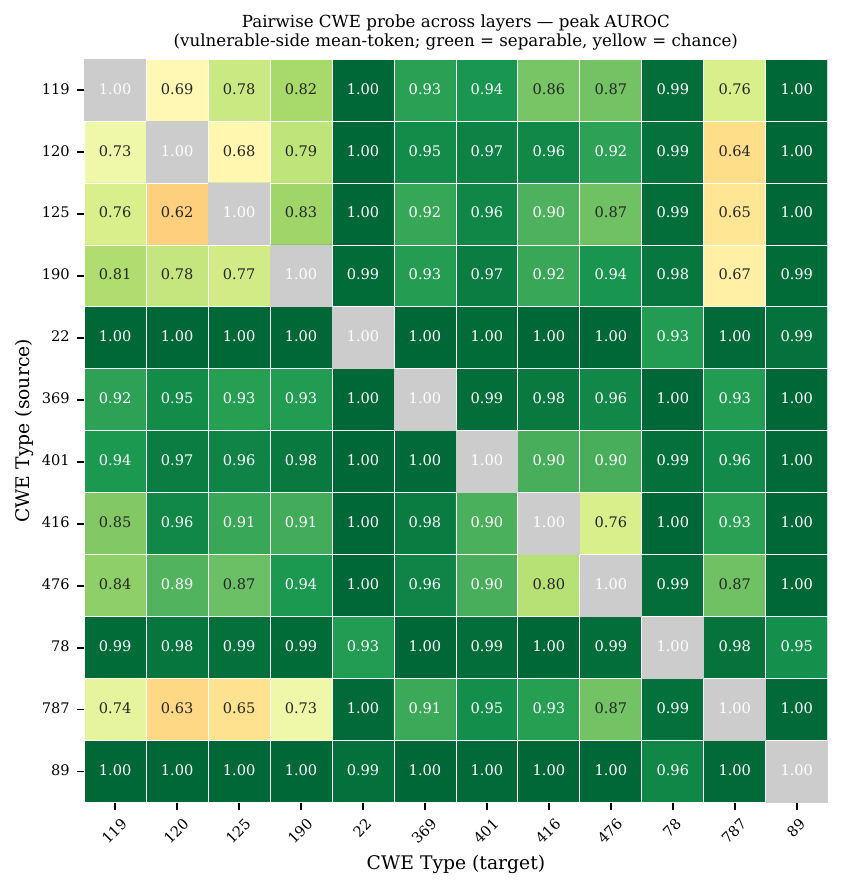}
    \caption{\textbf{Pairwise CWE-type probe AUROC.} Heatmap for \QwenFull on \DeltaSecommits (vulnerable-side mean-token pooling, Layer 19). The \ncwetypesdeltasecommits CWE types include memory violations (CWE-119, CWE-120, CWE-125, CWE-787), injection attacks (CWE-78, CWE-89), path traversal (CWE-22), and resource leaks (CWE-401). Most pairs are highly separable (90--100\% AUROC, dark green), with weaker separation between similar memory-related types (58--72\% AUROC, yellow/orange). Color scale: green indicates strong separation, yellow indicates chance-level performance. This pattern reflects strong functional encoding of vulnerability categories across diverse CWE types.}
    \label{fig:cwe_pairwise_qwen}
\end{figure}

We run pairwise binary CWE-type probes by extracting residual stream activations from vulnerable samples across all token positions, applying PCA-50 projection, and training a logistic regression classifier in 5-fold cross-validation. Figure~\ref{fig:cwe_pairwise_qwen} shows AUROC matrices for \DeltaSecommits.

\paragraph{Results: Strong CWE-type separation.}
CWE-type probing shows robust functional encoding: across the eight primary vulnerability categories shown in Figure~\ref{fig:cwe_pairwise_qwen}, most pairs exceed AUROC 0.85.
% , with mean pairwise AUROC of 91.5\% for \QwenFull, 91.9\% for \CodeLlamaFull, and 91.7\% for \StarCoderFull. 
This pattern of strong functional CWE separation replicates consistently across all \nmodels models spanning 6.7B to 15B parameters: \QwenFull and \QwenFourteenBFull, \CodeLlamaFull and \CodeLlamaThirteenBFull, \StarCoderFull and \StarCoderFifteenBFull, \DeepSeekFull, and \QwenCoderFull (see Appendix~\ref{app:multimodel} for per-model heatmaps). Mean pairwise AUROC across the 7B models averages 91.4\%, while larger models (13B--15B) average 90.8\%, indicating that vulnerability-type encoding is consistent across the tested scales (6.7B--15B). We do not claim scale invariance beyond this range: whether substantially larger models reorganize this signal remains an open question.

\paragraph{The encoding asymmetry.}
The model encodes distinct CWE-type patterns (buffer overflows, injection attacks, path traversal, resource leaks) with median pairwise AUROC 0.877. Vulnerable-vs.-secure classification achieves only peak AUROC 0.648 with mean-token pooling, using the same aggregation, probe, representation, and language. The +0.23 AUROC gap reveals that CWE-type patterns show higher linear separability, while vulnerability status (vulnerable/secure) shows lower categorical accuracy under these settings. 

\section{How Vulnerability Is Encoded}
\label{sec:results}

Section~\ref{sec:ablation} showed the model achieves high fidelity encoding \emph{what code does}, but achieves limited categorical accuracy on \emph{whether it is safe}. We now ask: where in the model's representations does this safety signal live? We find a consistent pattern across \nmodels code models spanning three model families and two size categories and three vulnerability datasets: the \VulnSignal exists and is geometrically stable, and---under our paired C-code datasets and linear-probing setup---is more accessible as a paired relational signal than as a globally separable categorical label.

\subsection{Consistency}
\begin{figure}[!htbp]
    \centering
    \includegraphics[width=\linewidth]{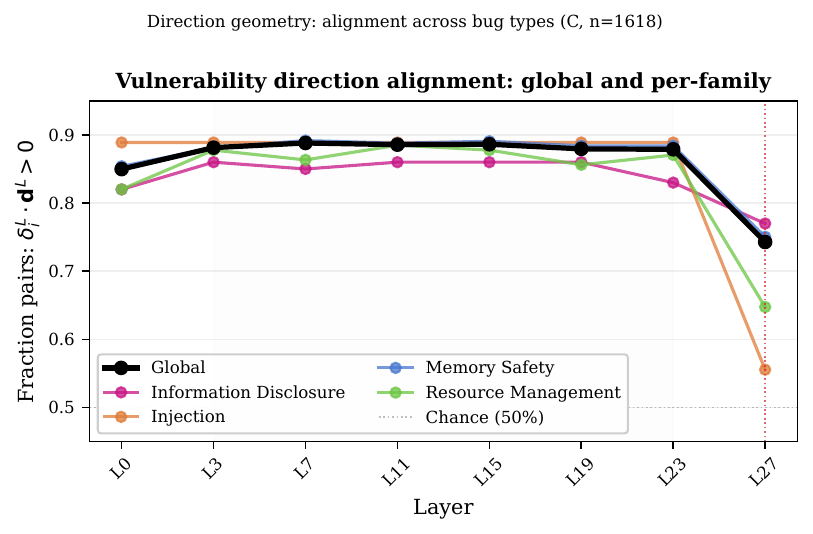}
    \caption{Vulnerability direction alignment across layers and bug types. Global alignment (black, bold) reaches 87--88\% across L3--L23, dropping to 70\% at L27. Per-family breakdown (colored lines) shows consistent alignment across bug types: Memory Safety (blue), Injection (orange), Resource Management (green), Information Disclosure (magenta). Family-level alignment closely tracks global alignment across all bug types.}
    \label{fig:direction_alignment}
\end{figure}

\paragraph{Multi-model Vulnerability direction consistency}
We find a stable, high-alignment \VulnDirection---replicates across all \nmodels models and all three vulnerability datasets. \SVEN produces alignment curves (86--87\% peak, 70\% at L27) and \PreciseBugs shows the same geometry. All three datasets show alignment profiles within 2 percentage points across all three models. Cross-model consistency is high (mean Pearson $r=0.921$ across the three model pairs), indicating that per-pair alignment values are nearly identical across the three independently trained architectures.

\textbf{(1) Geometric stability across layers.} The vulnerability direction is geometrically stable: cosine similarity between directions at different layers reaches 0.988--0.996 for all L3--L23 pairs (mean: 0.993, $n=15$ pairwise comparisons). The direction remains essentially identical until the final layer L27 (cosine similarity drops to $\approx 0$).

\textbf{(2) High per-pair consistency.} For 87--88\% of all vulnerable--secure pairs, the vulnerable representation lies further along the vulnerability direction than its secure counterpart (vs. 50\% baseline for random alignment). This alignment is stable from L3 to L23, dropping to 70.19\% at L27, reflecting the AUROC collapse at the final layer. Stability persists within L3--L23—showing vulnerability geometry emerges early and persists until the final layer, with 87--88\% representing a 74\% relative improvement over chance.

\begin{figure}[!htbp]
    \centering
    \includegraphics[width=\linewidth]{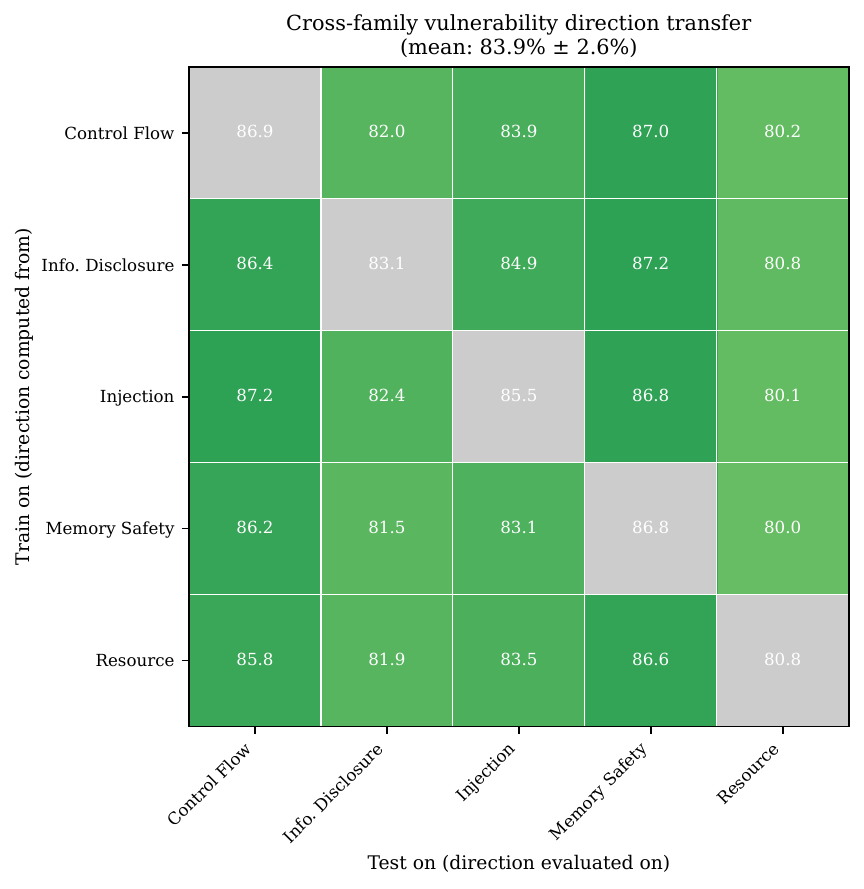}
    \caption{Cross-family vulnerability direction transfer heatmap (\QwenFull, \DeltaSecommits C subset, aggregated across layers L0--L27). Each cell shows per-pair alignment when a direction trained on the row family (on training data) is tested on the column family (on held-out test data), computed per-fold to maintain train/test separation. Values represent mean transfer across all layers. Warm colors (82--87\%) indicate strong transfer; cooler colors (70--75\%) indicate weaker transfer for injection attacks. The diagonal is undefined (comparing direction to itself). Mean cross-family transfer: 83.9\%, well above chance (50\%).}
    \label{fig:direction_transfer}
\end{figure}

\textbf{(3) Cross-family transfer.} We compute separate directions for memory-safety, injection, resource leaks, and information disclosure, then test whether directions trained on one family align with pairs from other families (Figure~\ref{fig:direction_transfer}). A direction learned from buffer-overflow family aligns with 83.9\% of information-disclosure pairs and 80.0\% of resource-leak pairs (vs. 50\% chance). Mean transfer across all families: 83.9\%, with memory-safety and information-disclosure at 81--87\%. Injection attacks show weaker cross-family alignment (70--75\%). This cross-family transfer indicates that directions are not entirely family-specific.

\paragraph{Generalization across model scales and families.}

The relational signal is consistent across the tested scales (6.7B--15B). Mean per-pair alignment (L3--L23) is nearly identical across model sizes: 7B models (\QwenFull, \CodeLlamaFull, \StarCoderFull) at 85.85\%, 13B--15B models (\QwenFourteenBFull, \CodeLlamaThirteenBFull, \StarCoderFifteenBFull) at 85.58\%, and alternative 7B models (\DeepSeekFull, \QwenCoderFull) at 83.23\%.

\subsection{Relational vs. categorical encoding}
\textbf{Why does this signal not support categorical classification with linear probes?} The direction captures \emph{paired comparisons}—in 87--88\% of pairs, the vulnerable code projects further along $\hat{d}$ than the secure code. However, in the full population, vulnerable and secure samples form overlapping distributions in projection space. This means no linear threshold on the identified vulnerability direction can separate them globally. The vulnerability information is encoded as relational structure—relative orderings between code pairs—rather than as categorical clusters, which limits the effectiveness of standard linear classification in absolute terms.

\begin{figure*}[!htbp]
    \centering
    \includegraphics[width=\linewidth]{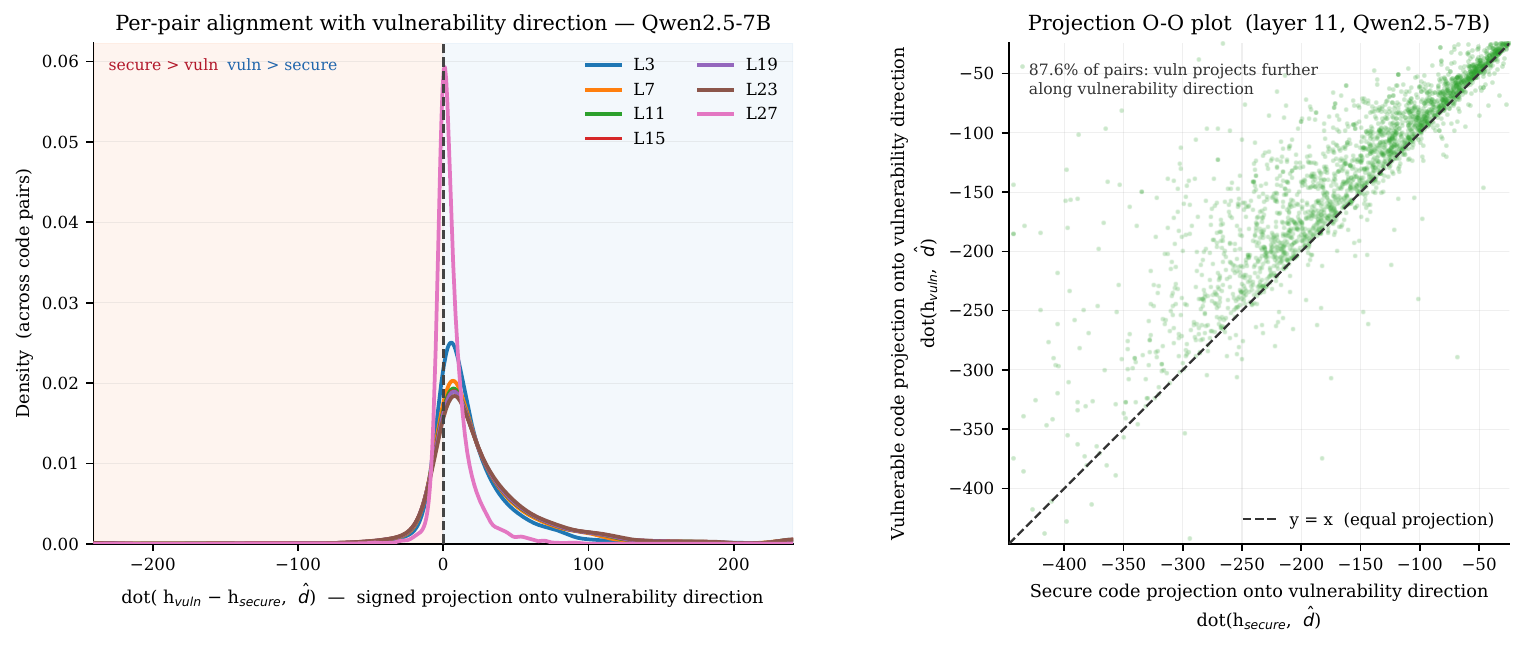}
    % \caption{Projection distributions reveal statistical regularity. \textbf{Left panel}: Histogram of pair-wise differences $(h_\text{vuln} - h_\text{secure}) \cdot \hat{d}$ for all pairs across layers (\QwenFull, mean-token pooling). The distribution is sharply peaked near zero with heavy tails—for most pairs. The mass concentrates close to zero rather than at large magnitudes. The remaining ~13\% of pairs are negative. \textbf{Right panel}: Scatter plot of how far each code version (vulnerable vs.\ secure) projects onto the direction. Points above the diagonal indicate vulnerable code projects further along the direction. The tight clustering around the diagonal demonstrates that while per-pair separation exists, individual pair-level projections are modest in magnitude. Vulnerable and secure samples form overlapping distributions in projection space, with no linear threshold that separates them globally.}
    \caption{Projection distributions (\QwenFull, mean-token pooling). \textbf{Left}: Distribution of pair-wise projection differences $(h_\text{vuln} - h_\text{secure}) \cdot \hat{d}$ across layers. Most pairs exhibit small positive differences concentrated near zero, with \(\sim\)13\% negative. \textbf{Right}: Vulnerable versus secure projections. The tight clustering around the diagonal and strong overlap between distributions indicate modest per-pair separation and no global linear decision~boundary.}
    \label{fig:projection_hist}
\end{figure*}

\textbf{A representation-level constraint in the tested settings:} This overlap persists across all models. Instead of globally separable clusters, the knowledge is encoded as pairwise orderings---``code A is more vulnerable than code B''---which is structurally distinct from absolute categorical labels---``code X is vulnerable.''

% This finding has a direct practical implication: 
Under standard linear probing the model achieves only modest absolute classification, but reliably ranks which code is riskier. This makes the model well-suited for prioritization and code-review triage, where relative risk matters more than absolute thresholds.

\paragraph{What each metric does and does not imply.} The two metrics answer different questions. Per-pair alignment is a \emph{relative} measure: given a matched (vulnerable, secure) pair, it reports whether the vulnerable version projects further along $\hat{d}$ than its secure counterpart. It presupposes a paired reference and quantifies ordering \emph{within} that reference. The binary probe is an \emph{absolute} measure: it asks whether a single, isolated representation can be labeled vulnerable or secure with no counterpart available. High per-pair alignment (87--88\%) therefore establishes that the direction encodes relative risk, but it does \emph{not} imply deployable vulnerability detection. In a realistic setting the model sees isolated code with no matched-secure reference, and absolute classification on the same direction is near chance (mean-token AUROC $\approx 0.648$); an explicit unpaired classification setting confirms this directly (peak AUROC $0.634$--$0.648$; Appendix~\ref{app:unpaired_probing}). Our alignment result should be read as evidence about representational structure, not as a detection score. Converting it into deployable detection requires supplying the missing reference---either a paired counterpart or a calibrated comparison set against which a sample's projection can be scored. Code review offers a natural instance of such a reference: a proposed change supplies its own before/after pair, so the direction can be applied as a lightweight helper process that flags whether an edit moves the representation toward the safer or the riskier side. In this pre-/post-change setting the relational signal is directly actionable---ranking or gating changes by the sign and magnitude of their projection shift---without requiring the absolute, per-sample classification that our linear probes do not recover.

\paragraph{Ruling out length artifacts.}
Vulnerable code is ~20\% longer on average in \DeltaSecommits, raising the question: could the vulnerability direction reflect sequence length rather than vulnerability properties? To test whether length is a confound, we employ a residualization approach: for each layer, we fit a regression model (Ridge with cross-validation) to predict activations from token count separately within vulnerable and secure samples. We then recompute the vulnerability direction on the residualized activations (actual minus length-predicted component). If the direction persists after length residualization, length differences do not fully explain the direction. This control is important because learned models can absorb length correlations; showing the direction survives length adjustment indicates the signal is not entirely length-driven. 
A per-model, per-layer breakdown confirms this: after regressing out log character count, the peak binary-probe AUROC remains $0.64$--$0.66$ across all \nmodels models, essentially unchanged from the uncontrolled peak of $0.648$ (Table~\ref{tab:length_residualized}).

\begin{table*}[!htbp]
\centering
\caption{\textbf{Length-controlled binary.} Vulnerable vs.\ secure probe AUROC with 95\% bootstrap CIs across layers, for all \nmodels models. Code length is regressed out before mean-token probing. Bold marks each model's peak layer. 
% Peak AUROC ($0.64$--$0.66$) is essentially unchanged from the uncontrolled value ($0.648$), confirming the signal is not a length artifact.
}
\label{tab:length_residualized}
\resizebox{\textwidth}{!}{%
\begin{tabular}{lccccccc}
\toprule
Model & L3 & L7 & L11 & L15 & L19 & L23 & L27 \\
\midrule
\CodeLlamaFull      & $0.610$ {\scriptsize $[0.592, 0.628]$} & $0.626$ {\scriptsize $[0.609, 0.644]$} & $\mathbf{0.644}$ {\scriptsize $[0.626, 0.659]$} & $0.640$ {\scriptsize $[0.623, 0.656]$} & $0.626$ {\scriptsize $[0.608, 0.643]$} & $0.598$ {\scriptsize $[0.580, 0.615]$} & $0.578$ {\scriptsize $[0.558, 0.594]$} \\
\CodeLlamaThirteenBFull & $0.611$ {\scriptsize $[0.592, 0.628]$} & $0.640$ {\scriptsize $[0.623, 0.656]$} & $\mathbf{0.646}$ {\scriptsize $[0.629, 0.663]$} & $0.644$ {\scriptsize $[0.627, 0.661]$} & $0.634$ {\scriptsize $[0.616, 0.650]$} & $0.618$ {\scriptsize $[0.600, 0.635]$} & $0.590$ {\scriptsize $[0.572, 0.608]$} \\
\DeepSeekFull       & $0.610$ {\scriptsize $[0.591, 0.628]$} & $0.631$ {\scriptsize $[0.614, 0.648]$} & $\mathbf{0.643}$ {\scriptsize $[0.625, 0.659]$} & $0.635$ {\scriptsize $[0.618, 0.653]$} & $0.630$ {\scriptsize $[0.612, 0.647]$} & $0.614$ {\scriptsize $[0.595, 0.631]$} & $0.607$ {\scriptsize $[0.590, 0.624]$} \\
\QwenFull           & $0.613$ {\scriptsize $[0.595, 0.632]$} & $0.629$ {\scriptsize $[0.611, 0.646]$} & $0.628$ {\scriptsize $[0.610, 0.645]$} & $0.645$ {\scriptsize $[0.626, 0.661]$} & $\mathbf{0.654}$ {\scriptsize $[0.637, 0.670]$} & $0.623$ {\scriptsize $[0.605, 0.640]$} & $0.604$ {\scriptsize $[0.585, 0.622]$} \\
\QwenCoderFull      & $0.603$ {\scriptsize $[0.584, 0.622]$} & $0.627$ {\scriptsize $[0.609, 0.644]$} & $0.620$ {\scriptsize $[0.602, 0.637]$} & $0.648$ {\scriptsize $[0.629, 0.665]$} & $\mathbf{0.657}$ {\scriptsize $[0.640, 0.674]$} & $0.614$ {\scriptsize $[0.595, 0.631]$} & $0.585$ {\scriptsize $[0.566, 0.603]$} \\
\QwenFourteenBFull  & $0.610$ {\scriptsize $[0.593, 0.629]$} & $0.625$ {\scriptsize $[0.607, 0.643]$} & $0.608$ {\scriptsize $[0.589, 0.626]$} & $0.620$ {\scriptsize $[0.602, 0.637]$} & $0.631$ {\scriptsize $[0.614, 0.648]$} & $0.643$ {\scriptsize $[0.625, 0.659]$} & $\mathbf{0.646}$ {\scriptsize $[0.628, 0.662]$} \\
\StarCoderFull      & $0.625$ {\scriptsize $[0.607, 0.643]$} & $\mathbf{0.644}$ {\scriptsize $[0.625, 0.661]$} & $0.637$ {\scriptsize $[0.620, 0.654]$} & $0.634$ {\scriptsize $[0.618, 0.652]$} & $0.628$ {\scriptsize $[0.611, 0.645]$} & $0.610$ {\scriptsize $[0.593, 0.627]$} & $0.596$ {\scriptsize $[0.578, 0.613]$} \\
\StarCoderFifteenBFull & $0.617$ {\scriptsize $[0.599, 0.636]$} & $\mathbf{0.641}$ {\scriptsize $[0.623, 0.658]$} & $0.632$ {\scriptsize $[0.615, 0.649]$} & $0.624$ {\scriptsize $[0.607, 0.641]$} & $0.634$ {\scriptsize $[0.616, 0.650]$} & $0.631$ {\scriptsize $[0.613, 0.648]$} & $0.606$ {\scriptsize $[0.588, 0.623]$} \\
\bottomrule
\end{tabular}%
}
\end{table*}

\section{Discussion}
\label{sec:discussion}

\paragraph{Implications for vulnerability detection.}
Across all \nmodels models studied, and under the paired C-code datasets and linear-probing setup we consider, vulnerability information is more accessible as a paired relational signal than as a globally separable categorical label. Representation-based categorical classification via linear probes on standard pooling is insufficient for standalone deployment; ranking-based and preference-learning approaches align with how vulnerability is encoded in the models tested. Recent work on constrained decoding~\cite{fu2024constrained} and secure code generation~\cite{zhang2024seccoder} demonstrates the effectiveness of methods that leverage ordinal relationships in code representations.

\paragraph{Implications for representation structure in code generation.}
Our conclusion is not that absolute vulnerability classification is impossible, but that it is weakly recoverable under our linear probing setup. A shallow MLP probe on the same representations (64 hidden units; Appendix~\ref{app:mlp_probe}) yields peak AUROC $0.635$--$0.644$ across models, essentially unchanged from the linear-probe peak of \AUROCPeak, indicating that a lightweight nonlinear readout does not recover materially more categorical signal. We do not preclude that more expressive readouts recover stronger absolute performance---larger nonlinear probes, learned or token-level pooling, contrastive objectives, or prompt-based classifiers with richer natural-language context. Rather, our findings characterize what is accessible with linear and lightweight nonlinear readouts on standard pooling: the model reliably ranks which code is riskier, but categorical classification achieves limited accuracy due to overlapping distributions. Whether more expressive nonlinear methods can recover higher categorical accuracy on this representation remains an open question orthogonal to characterizing the representation's structure. Our analysis identifies a single global vulnerability direction with strong geometric properties: stability across layers, consistency across \nmodels model architectures and scales, and transferability across bug families. These consistent properties demonstrate a unified relational vulnerability signal in these models. However, whether a single global direction suffices for all vulnerability types and contexts, or whether mixtures/local directions could better capture context-dependent vulnerability semantics, remains an empirical question. Recent representation-engineering work such as \citet{yu2025moc} (mixtures of control vectors), \citet{steer_your_model}, and \citet{yousefpour2025representation} has explored leveraging learned directions to influence model behavior. Systematic evaluation of whether single vs. mixture directional approaches are most effective for steering on the identified vulnerability direction is future work. Related approaches like \citet{adila2026weightupdatesactivationshifts} and \citet{belitsky2025kvcachesteeringcontrolling} propose complementary mechanisms (weight updates, KV cache interventions) for controlled manipulation.

\paragraph{Limitations and scaling implications.}
Our conclusions are scoped to the studied setting and should not be read as claims about code models or LLMs in general.
(1)~Model coverage spans \nmodels open models in the 6.7B--15B range; whether the same representation geometry holds for substantially larger or closed frontier models is untested.
(2)~The analysis is intentionally restricted to C programs to control for language-specific variation while isolating representational properties, so transfer to other programming languages (Python, Java, JavaScript) is unverified; evaluating additional languages is an important direction for future work.
(3)~We probe fixed, context-free representations; instruction-following vulnerability detection with richer natural-language context is outside our scope and may expose additional categorical structure.
(4)~Reported AUROC values depend critically on two factors. Cross-validation methodology (pair-stratified vs.\ standard fold; see Methodology~\ref{sec:stats}) contributes +0.055 for last-token, +0.016 for mean-token. Token aggregation strategy contributes +0.117.
(5)~Paired vulnerable/patched datasets naturally encode a before/after relationship, so the learned direction may capture general code-change signals in addition to vulnerability-specific patterns. Disentangling the two would require comparable paired data from other software-engineering activities such as refactoring or code simplification. Such data remain scarce for C: \citet{ritz2025refactoring} introduce a C++ refactoring dataset with only 16 examples, and building reliable resources would require mining commits explicitly marked as refactoring together with robust validation. We leave this investigation to future work.

\paragraph{Reproducibility and hyperparameter details.}
To support future replication and mechanistic interpretability work, we intend to release all code, activation caches, and analysis scripts upon publication. \textbf{Probing:} Linear logistic regression, L2 regularization ($\lambda = 1.0$), balanced class weighting. PCA-50 projection (61.3\% variance), fitted independently per fold. PCA-100 differs <0.01 (Appendix~\ref{app:pca_sensitivity}). \textbf{Direction computation:} Vulnerability direction $\mathbf{d} = \mathrm{normalize}(\bar{v} - \bar{s})$ per layer, using mean activations from training data.

Advances in the field have hindered the practical adoption of LLMs in software vulnerability detection, including limited robustness in code representations, vulnerability data leakage, and limited interpretability~\cite{asystematicliteraturereviewonllmsforvulnerabilitydetection, TowardsExplainableVulnerability,llmsinsoftwaresecurity, zheng2023survey, ramirez2024state}. We hope our findings on the representational structure of vulnerability signals can inform future model design and future training strategies to address these limitations, particularly through comprehensive approaches~\cite{dai2025comprehensive} that account for the greater accessibility of vulnerability information as a relational signal than as a globally separable categorical label in these settings.

The consistent direction vectors across layers align with recent work on representation engineering (\citet{zou2023repeng}), particularly approaches like \citet{yu2025moc} which leverage collections of linear directions to steer code generation. Recent work on contrastive learning and preference modeling are naturally aligned with relational vulnerability signals, given that these methods directly optimize pairwise comparisons. Empirical validation of steering and preference-learning approaches on this direction remains future work. Complementary approaches such as integrated static analysis and fuzz testing~\cite{li2024enhancing}, combined with knowledge-level retrieval-augmented generation~\cite{du2024vul}, represent strategies for enhancing LLM-based vulnerability detection by combining relational signals with additional safety mechanisms. Broader strategies including graph-based vulnerability detection~\cite{lu2024grace}, dataset-driven vulnerability mitigation~\cite{wang2023enhancing}, similarity-based embeddings~\cite{shimmi2024vulsim}, and systematic evaluation frameworks~\cite{peng2025cweval} demonstrate the diversity of approaches complementary to relational encoding.

Our work is a representation analysis, not an end-to-end %vulnerability detection 
system. Recent work shows that richer context and structured reasoning substantially improve absolute detection. 
Those findings are orthogonal to and consistent with ours, since we characterize how vulnerability is organized in fixed, context-free representations. The relative encoding we observe is best exploited by ranking-based retrieval and human-in-the-loop triage---comparing the relative risk of code samples rather than performing direct binary classification. Detection systems could combine this relational signal with additional context and reasoning mechanisms.

\section{Conclusion}

The code models we study understand vulnerability, but encode it in a form that differs from how we typically deploy them. They reliably rank pairs of code snippets by risk (87–88\% per-pair alignment) yet achieve only moderate categorical classification accuracy (peak AUROC 0.648). Under the studied paired C-code datasets and linear probing, the \VulnSignal is more accessible as relational structure—comparative judgments like ``this is worse than that''—than as a globally separable categorical label.

\section*{Disclosure}

AI was used to improve the clarity and organization of this manuscript.

\section*{Acknowledgments}

% This work was supported by Fundação para a Ciência e Tecnologia (FCT): through the CMU Portugal Dual PhD program: 
Rui Melo (2024.18549.PRT) is funded by Fundação para a Ciência e Tecnologia (FCT)
% Rui Melo (2024.18549.PRT), Claudia  Mamede (PRT/BD/155043/2022), and André Catarino.
This work is also supported by the SALVE project (2024.14936.PEX) and HPC (2025.08039.CPCA).

\section*{Societal Impact}

Our work is motivated by a defensive objective: understanding how code LLMs internally represent vulnerability-related information to support human-in-the-loop vulnerability triage and safer code-generation systems.
We acknowledge a dual-use risk: a stable vulnerability direction could reduce the cost of prioritizing code for security inspection. Although our techniques build on existing representation analysis methods, they may lower the barrier to misuse. We therefore present this work as a resource for defensive evaluation and understanding model behavior, rather than as an operational attack toolkit.
\clearpage
\bibliography{biblio}

\end{document}